\documentclass[%
aps, 
prb,
twocolumn, oneside, 10pt, amsfonts, amsmath, nofootinbib, floatfix, superscriptaddress] {revtex4-2}

\usepackage{graphicx}
\usepackage[colorlinks=true]{hyperref}
\usepackage{textcomp}
\usepackage{physics}

\newcommand{\rv}{{\bf r}}
\newcommand{\Rv}{{\bf R}}
\newcommand{\kv}{{\bf k}}
\newcommand{\qv}{{\bf q}}
\newcommand{\NPR}{\textrm{NPR}}
\newcommand{\CL}{{\cal L}}
\newcommand{\CT}{{\cal T}}
\newcommand{\lc}{{\lambda_\textrm{C}}}
\newcommand{\ec}{{E_\textrm{C}}}

\begin{document}
\title{Emergence of quasiperiodic Bloch wave functions in quasicrystals}

\author{Omri Lesser}
\affiliation{Department of Condensed Matter Physics, Weizmann Institute of Science, Rehovot 7610001, Israel}
\author{Ron Lifshitz}
\email[Corresponding author:\ ]{ronlif@tau.ac.il}
\affiliation{Raymond and Beverly Sackler School of Physics \& Astronomy, Tel Aviv University, Tel Aviv 69978, Israel}
\affiliation{School of Mathematics, University of Leeds, Leeds LS2 9JT, United Kingdom}

\date{\today}

\begin{abstract}
We study the emergence of quasiperiodic Bloch wave functions in quasicrystals, employing the one-dimensional Fibonacci model as a test case. We find that despite the fact that Bloch functions are not eigenfunctions themselves, superpositions of relatively small numbers of nearly degenerate eigenfunctions give rise to extended quasiperiodic Bloch functions. These functions possess the structure of earlier ancestors of the underlying Fibonacci potential, and it is often possible to obtain different ancestors as different superpositions around the same energy. There exists an effective crystal momentum that characterizes these ancestors, which is determined by the mean energy of the superimposed eigenfunctions, giving rise to an effective dispersion curve. We also find that quasiperiodic Bloch functions do emerge as eigenfunctions when weak disorder is introduced into the otherwise perfect quasiperiodic potential. These theoretical results may explain a number of experimental observations, and may have practical consequences on emerging theories of band topology and correlated electrons in quasicrystals.
\end{abstract}


\maketitle

\section{Introduction and motivation}

Quasicrystals are quasiperiodic crystals that are strictly aperiodic~\cite{[{For definitions and terminology see, for example,\ }] crystaldef, *quasidef}. Their discovery~\cite{Shechtman84} put an end to the age-old paradigm that long-range order---in the positions of atoms in a material---is synonymous with periodicity, ushering in a Kuhnian scientific revolution~\cite{Cahn96}. In the past four decades we have experienced the exciting paradigm-rebuilding phase of this revolution~\cite{rebirth}, in which old notions and ideas are being reexamined, and carefully modified and adapted to the age of quasicrystals~\cite{symbreak}. 
Of particular relevance here is the fundamental notion that order, or lack thereof, is what determines the nature of single-particle quantum excitations in condensed matter. 

Bloch~\cite{bloch} showed, already in 1928, that electrons in periodic potentials form extended eigenfunctions, almost as if they were in free space, with absolutely continuous energy spectra. Anderson~\cite{Anderson58} established three decades later that electrons in disordered potentials possess exponentially localized eigenfunctions, with pure point spectra. Surprisingly, only in dimensions greater than two is there a delocalization transition as a function of energy~\cite{Abrahams79}, whereby beyond a critical value---the so-called mobility edge---electrons have sufficient energy to overcome the disorder for their eigenfunctions to become extended. It was only natural for the scientific community, which at the time was still unaware of aperiodically ordered matter, to associate the existence of extended or localized eigenfunctions with having ordered or disordered potentials, respectively. The discovery of quasicrystals called for a reexamination of this notion.

Quasicrystals possess long-range order. It is therefore natural to expect their single-particle eigenfunctions to be extended and obey a generalized form of Bloch's theorem for quasiperiodic potentials. Yet, four decades of research on single-particle excitations in quasicrystals~\cite{Dinaburg, Russman80, frohlich90, Eliasson92, AAmodel, Sinai87, Kohmoto83, *Kohmoto86, *Kohmoto87, ostlund, Niu86, *Niu90, Ashraff88, KKL, Luck86, Lu87, Azbel79, Kollar86, Bellissard82, Holzer88, Wurtz88, Ghosh91, Baake92, Bovier95} have shown that this is not the case. Depending on the nature of the model used, single-particle eigenfunctions in quasicrystals can take almost any possible form: they can be extended as in periodic potentials, localized as in disordered potentials, or critical, exhibiting power-law or algebraic decay, which is not seen in the former cases.

As it seems, the Bloch theorem, which is the cornerstone of the theory of electrons in periodic crystals, does not generally apply to aperiodic crystals. Rather than proposing alternative forms for the electronic eigenfunctions in quasicrystals~\cite{Sutherland86, Kalugin14}, or suggesting a generalized formulation for the Bloch theorem~\cite{Akkermans21}, we examine here whether Bloch wave functions may still appear naturally in quasicrystals even when Bloch's theorem fails to generate them as eigenfunctions. 
This may have immediate practical implications, as the study of correlated electrons in quasicrystals~\cite{Vidal99,Hida01} is regaining interest~\cite{Takemori15a, *Takemori15b, *Takemura15, *Takemori20, Araujo19, roy_reentrant_2021}, amid the experimental observation of Dirac electrons in dodecagonal bilayer graphene quasicrystals~\cite{Ahn18,Yan19}, and the reformulation of the early spectral theories of electrons, phonons, and photons in quasicrystals~\cite{Sire89, Bellisard92a, *Bellisard92b} using modern topological classifications~\cite{Kraus12a, *Kraus12b, *Kraus13, Madsen13, Bandres16, Dareau17, li_mobility_2020}. It may also have experimental significance with the increasing ability, at least in metamaterials, to observe and manipulate the actual wave functions~\cite{Lahini09} and their topological nature~\cite{Verbin13, *Verbin15, Tanese14, *Baboux17, *Goblot20, an_interactions_2021}, while inducing topological defects~\cite{Freedman06}, phason excitations~\cite{Freedman07}, and static disorder~\cite{Levi11} in the otherwise perfectly ordered quasicrystalline potential.

To be concrete, let us consider a quasiperiodic crystalline potential $U(\rv)$, which can be decomposed into countably many Fourier modes, 
\begin{equation}\label{Eq:potential}
U(\rv) = \sum_{\kv \in \CL} \tilde{U}(\kv) e^{i\kv\cdot\rv},
\end{equation}
where the reciprocal lattice $\CL$, consisting of the closure under addition of all wave vectors $\kv$ with nonzero Fourier coefficients $\tilde{U}(\kv)$, is a finitely generated $\mathbb{Z}$-module. If the rank, or smallest number $D$ of wave vectors required to generate $\CL$ over the integers, is equal to the spatial dimension $d$, then $U(\rv)$ is the potential of a periodic crystal. In other words, if $D=d,$ $\CL$ is reciprocal to a Bravais lattice $\CT$ of real-space translations leaving the crystalline potential invariant, in the sense that $U(\rv+\Rv)=U(\rv)$ for every $\Rv\in \CT$. If $D>d$, then $U(\rv)$ is the potential of a quasicrystal, there exists no lattice of spatial translations leaving the potential invariant, and the reciprocal lattice $\CL$ becomes dense and no longer possesses a Brillouin zone.

According to the Bloch theorem (see, for example, chapter 8 of \citet{AM}), single-electron eigen\-functions, satisfying the Schr\"{o}dinger equation for such a potential, can be expressed as 
\begin{equation}\label{Eq:Bloch}
\psi_{\qv} (\rv) = e^{i\qv\cdot\rv} \sum_{\kv \in \CL} \tilde{u}_{\qv} (\kv) e^{i\kv\cdot\rv},
\end{equation}
where the quantum number $\qv$ labeling them is known as the crystal momentum.  Consequently, the Fourier transforms of the electron density $|\psi_{\qv} (\rv)|^2$ and the potential $U(\rv)$ are supported on the same reciprocal lattice $\CL$, while the Fourier transform~\eqref{Eq:Bloch} of the eigenfunction itself is supported on the shifted lattice $\CL+\qv=\left\{\kv+\qv\ |\ \kv\in \CL \right\}$.
By comparing the Fourier transforms of $|\psi_{\qv}(\rv)|$ and $\psi_{\qv} (\rv)$, one can readily determine the crystal momentum $\qv$ by the presence of a uniform shift between the two spectra. Repeating this process for all eigenfunctions yields a dispersion relation $E(\qv)$, which corresponds to the crystalline band structure. This holds true if the crystal is periodic. It generally fails if the crystal is aperiodic, possessing a dense reciprocal lattice, owing to the fact that the infinite sum over scattered waves in Eq.~\eqref{Eq:Bloch} generically does not converge, even though the corresponding sum in Eq.~\eqref{Eq:potential} does. 

Nevertheless, it is interesting to explore whether extended quasiperiodic Bloch functions may somehow emerge in realistic physical situations---either as naturally occuring superpositions of eigenfunctions, or even as true eigenfunctions of slightly modified quasiperiodic structures, for example, by introducing static disorder or external fields. If so, what would be the nature of these quasiperiodic Bloch functions? How would they be related to the structure of the underlying quasicrystalline potential? Could they be characterized by similar quantum numbers $\qv$ in reciprocal space as their periodic analogs? If so, would there exist an energy-momentum dispersion curve, or effective band structure, that could be associated with these Bloch functions? Positive answers to these questions may explain a number of empirical observations that seem to indicate that energy-momentum dispersion curves do exist~\cite{rotenberg,Chiang19}, and that slight disorder may increase the conductivity~\cite{Roche97, *Mayou00} and the spatial extent of wave functions~\cite{Levi11}. The latter effect has been demonstrated recently using renormalization-group calculations~\cite{Jagannathan19, *Jagannathan20}.

\section{Choice of model}

We would like to choose, among the commonly used models for studying electrons in quasicrystals, one that is as simple as possible, yet sufficiently generic to explore the questions raised above.
We therefore concentrate on one dimension---although some initial calculations of ours in two dimensions can be found elsewhere~\cite{ShaharArxiv, ShaharThesis}---and limit ourselves to quasicrystals of rank $D=2$, that is with just a single pair of incommensurate fundamental spatial harmonics. With these restrictions there are still a few different types of models to choose from, some more physically motivated than others. These can roughly be categorized as follows (with more details in the references provided):

\subsection{Continuous time-independent Schr\"{o}dinger equations}
\label{Sec:ModelA}

These are ordinary differential equations of the form
\begin{equation}\label{Eq:ContSchrodinger}
        -\psi''(x) + \lambda U(x) \psi(x) = E\psi(x),
        \quad x\in\mathbb{R},
\end{equation}
with continuous quasiperiodic potentials $U(x)$ as in Eq.~\eqref{Eq:potential}, supported on a reciprocal lattice $\cal L$ of rank $D=2$~\cite{Dinaburg, Russman80, frohlich90, Eliasson92}. To simplify things, one often limits the potential to its two fundamental harmonics by taking, for example,
\begin{equation}\label{Eq:SumofCosines}
        U(x) = \cos 2\pi x + \epsilon\cos 2\pi\left(\tau x + \theta\right),
\end{equation}
with an irrational $\tau$.

For weak coupling $\lambda\ll1$, or weak quasiperiodicity $\epsilon\ll1$, one can treat the equation as a small perturbation with respect to free electrons, or to Bloch electrons in a periodic crystal, respectively. In these limits the spectrum remains purely absolutely continuous, exhibiting a well-defined band structure $E(q)$, with a hierarchy of gaps that open at $q=n\pi + m\pi\tau$ ($n,m\in\mathbb{Z}$), as a result of the hybridization that occurs whenever the degenarate $\pm q$ free-electron states differ by a wave vector $k\in\cal L$. Formally, this happens because of small divisors that appear in the perturbation expansion (see, for example, chapter 9 of \citet{AM}). In these limits the eigenfunctions are quasiperiodic and satisfy Bloch's theorem~\eqref{Eq:Bloch}.

Owing to the dense nature of $\cal L$, for sufficiently large $\lambda$ and $\epsilon$, and depending on the Diophantine properties of the irrational ratio $\tau$, the formation of gaps typically destroys the band structure at the bottom of the spectrum. Consequently, the bottom of the spectrum becomes pure point with eigenfunctions that are exponentially localized. Typically~\cite{Yao19}, there is a critical energy $\ec = \ec(\lambda,\epsilon,\tau)$ below which the eigenfunctions are exponentially localized and above which they are extended, akin to the mobility edge that is observed in Anderson localization in dimensions greater than 2. It should be emphasized, though, that while Anderson localization arises from the lack of order, the localization here arises from the existence of order, albeit aperiodic order. It is the long-range order that is responsible for having strong Bragg peaks $\tilde{U}(k)$ in the Fourier transform of the potential~\eqref{Eq:potential}, that can then combine with the dense nature of $\cal L$ to destroy the continuous bands.

\subsection{Discrete time-independent Schr\"{o}dinger equations}
\label{Sec:ModelB}

These are finite difference equations of the form
\begin{equation}\label{Eq:AAmodel}
      \psi(m+1) + \psi(m-1) + \lambda f(m)\psi(m)
      = E\psi(m),
\end{equation}
with a periodic analytic function $f(x)$ whose period is incommensurate with that of the lattice $m\in\mathbb{Z}$ on which it is sampled. A standard example is to take 
\begin{equation}\label{Eq:AAFunction}
        f(x) = \cos 2\pi(\tau x + \theta),
\end{equation} 
with an irrational $\tau$. 

In these finite-difference, or tight-binding, Schr\"{o}dinger equations~\eqref{Eq:AAmodel}, known as the Aubry-Andr\'{e} model~\cite{AAmodel}, or the almost Mathieu equation~\cite{Sinai87, frohlich90, Last95, Gordon97, Jitomirskaya99}, one no longer observes a localization transition as a function of energy. Instead, what one typically finds is that there is a critical value $\lc$ of the coupling constant, equal to 2 for the standard example~\eqref{Eq:AAFunction}, at which the whole spectrum changes its nature~\cite{Jitomirskaya99}. For almost every choice of irrational $\tau$ and real $\theta$, if $\lambda<\lc$ the bands remain intact and the whole spectrum is purely absolutely continuous with eigenfunctions that are extended. If $\lambda>\lc$ the whole spectrum is pure point and the eigenfunctions are exponentially localized. Exactly at $\lambda=\lc$ the spectrum is purely singular continuous~\cite{Gordon97}, like a Cantor set---uncountable yet of zero Lebesgue measure---neither pure point nor consisting of any bands or continuous intervals. The corresponding eigenfunctions decay algebraically. 

\subsection{Tight-binding time-independent Schr\"{o}dinger equations on quasiperiodic tilings}
\label{Sec:ModelC}

In one dimension, quasiperiodic tilings
reduce to quasiperiodic sequences of finitely many different letters. These letters may then represent either the sequence of distances between atoms on the line, affecting the hopping amplitudes $\{T_m\}$ between neighbors, the sequence of different atomic species arranged on the line, affecting the onsite energies $\{\varepsilon_m\}$, or both, as given by
\begin{equation}\label{eq:tight-general}
    T_{m+1} \psi(m+1) + T_m \psi(m-1) + \varepsilon_m \psi(m)
      = E\psi(m).
\end{equation}
When the $T_m$ are all equal, or the $\varepsilon_m$ are all equal, Eq.~\eqref{eq:tight-general} reduces to the diagonal or the off-diagonal tight-binding models, respectively. 

The essential difference between these equations and those of the previous category is that the sequence $f(m)$ in Eq.~\eqref{Eq:AAmodel} consists of infinitely many different values, densely sampling the image of the continuous function $f(x)$, whereas the sequences $\{T_m\}$ or $\{\varepsilon_m\}$ consist of a finite number of distinct values. A standard example is the family of Sturmian sequences~\cite{Damanik99}, consisting of two letters, corresponding to whether vertical or horizontal lines are crossed when cutting through a square grid with an irrationally sloped straight line. This includes the most familiar example of the Fibonacci sequence~\cite{Kohmoto83, ostlund, Niu86, Ashraff88, KKL, Luck86}, recently reviewed by \citet{Jagannathan21}.

The striking behavior for a large family of these models~\cite{Damanik99, Jagannathan21, Damanik00, Suto89, Bellissard89} is that they exhibit purely singular-continuous zero-Lebesgue-measure spectra, with critical eigenfunctions, for \emph{any} nontrivial choice of their parameters. They never exhibit absolutely continuous spectra with extended Bloch eigenfunctions, as do periodic crystals, and they never exhibit pure point spectra with exponentially localized eigenfunctions, as disordered structures do. They represent what is most unique about quasicrystals in this respect, and are therefore best suited to study the main question posed here: Is it still possible for Bloch wave function to occur naturally in these aperiodically ordered models, despite the failure of Bloch's theorem in any part of their spectra and for any choice of their parameters.

\subsection{Time-independent quasiperiodic Kronig-Penney models}
\label{Sec:ModelD}

The last family of models that should be mentioned in this context are quasiperiodic generalizations of the 1931 Kronig-Penney model~\cite{Kronig31}. These models~\cite{Azbel79, Kollar86, Bellissard82, Holzer88, Wurtz88, Ghosh91, Baake92, Bovier95} are in some sense intermediate between the continuous models of Sec.~\ref{Sec:ModelA} and the discrete models of Secs.~\ref{Sec:ModelB} and~\ref{Sec:ModelC}, as they replace the continuous potential $U(x)$ in the Schr\"{o}dinger equation~\eqref{Eq:ContSchrodinger} by a discrete sum of delta functions with variable weights $W_m$,
\begin{equation}\label{Eq:KP}
        U(x) = \sum_{m\in\mathbb{Z}} W_m\delta(x-x_m).
\end{equation}
Taking $x_m=m$ and setting $W_m=f(m)$, gives the Kronig-Penney version of the almost Mathieu equation or Aubry-Andr\'{e} model of Eq.~\eqref{Eq:AAmodel}; while taking the distances $x_m-x_{m-1}$ or the weights $W_m$ to follow a quasiperiodic sequence of letters, gives the Kronig-Penney version of the tight-binding model of Eq.~\eqref{eq:tight-general}. Other variations are possible, including the replacement of the weights $W_m$ by a convolution of the delta functions with finite-range potentials $v_m(x)$---like square wells or barriers that are narrower than the minimum distance between consecutive delta functions.

These quasiperiodic Kronig-Penney models typically exhibit the same features as the discrete tight-binding versions, usually with some added complexities that are lost in the tight-binding approximation. An interesting example is the appearance of countably many extended Bloch eigenfunctions, whose energies are embedded within the singular continuous spectrum of an uncountable Cantor set of critically decaying eigenfunctions~\cite{Baake92, Bovier95}. 

\subsection{Our choice: The off-diagonal tight-binding Fibonacci model}

As argued in Sec.~\ref{Sec:ModelC} above, we choose to consider the family of tight-binding models of Eq.~\eqref{eq:tight-general}, where the Bloch theorem fails completely, and the eigenfunctions are all critically decaying for any nontrivial choice of parameters. We employ the well-known Fibonacci sequence, taking the off-diagonal version of the model for simplicity,
\begin{equation}\label{eq:tight}
  T_{m+1} \psi(m+1) + T_m \psi(m-1)
  = E\psi(m).
\end{equation}
Here, the onsite energies are all zero, and the hopping amplitudes $T_m$, between pairs of adjacent sites $\psi(m)$ and \mbox{$\psi(m-1)$}, take the values $1$ or $T$, depending on whether the two sites are separated by a Long ($L$) or a Short ($S$) interval, respectively. The two types of intervals are arranged according to the infinite Fibonacci sequence $S_\infty = \lbrace L, S, L, L, S, L, S, L, L, S, L, L, S, \ldots\rbrace$, which can be obtained, for example, as the infinite limit of a recursive concatenation of finite sequences, whereby $S_n=\{S_{n-1}S_{n-2}\}$, starting with $S_1=\{S\}$ and $S_2=\{L\}$. Accordingly, the length of the finite sequence $S_n$ is the Fibonacci number $F_n$, where $F_n=F_{n-1}+F_{n-2}$ and $F_1=F_2=1$. This is the same as starting with the sequence $S_1=\{S\}$ and iteratively applying the substitution rules $S\to L$ and $L\to LS$. Moreover, the finite sequences $S_n$ yield the optimal choices of unit cells for periodic approximants of the infinite Fibonacci quasicrystal. A finite sequence $S_m$ is said to be an \emph{ancestor} of the sequence $S_n$ if $m<n$. 

\begin{figure}
\centering
\includegraphics[width=\linewidth]{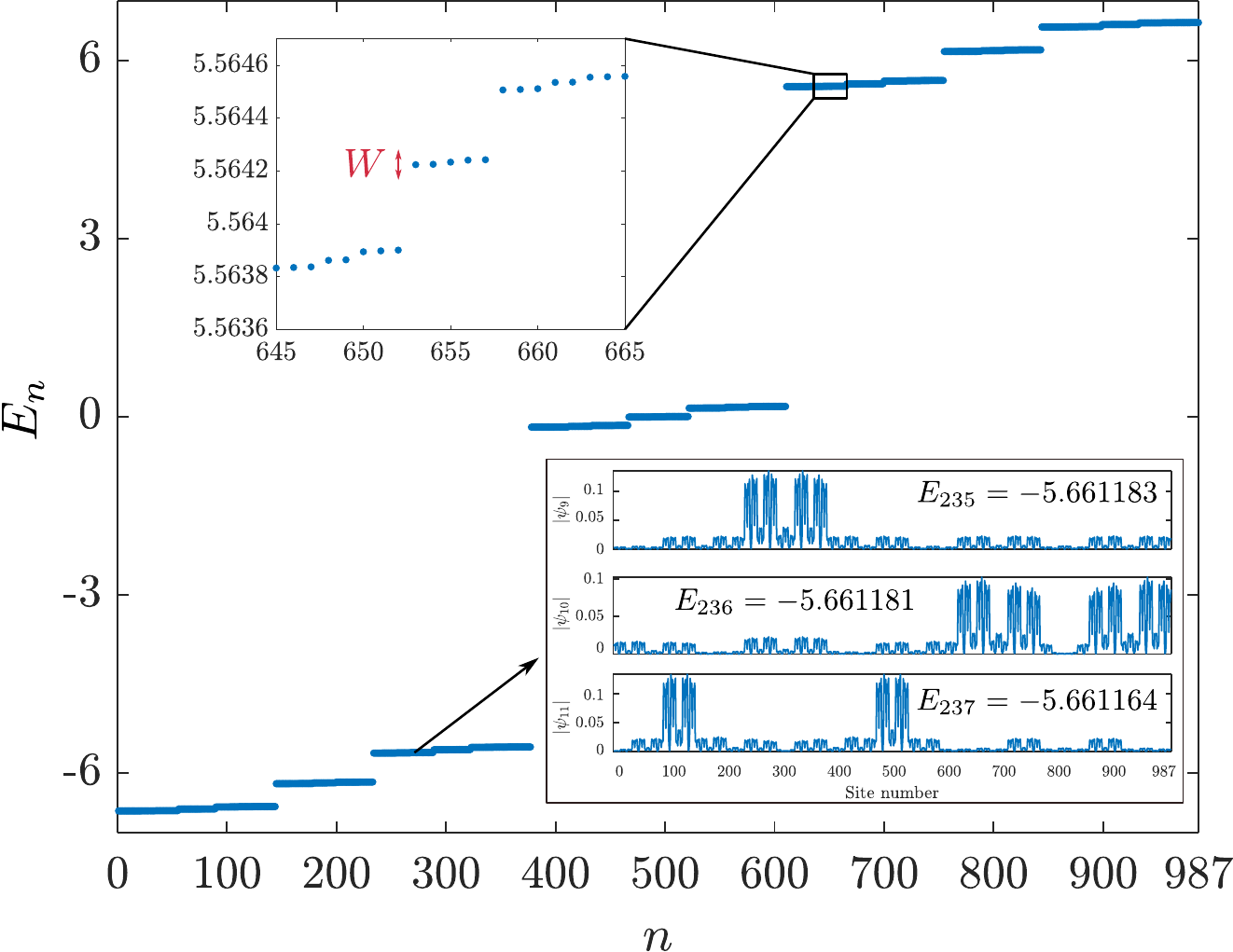}
\caption{Energy spectrum of the off-diagonal tight-binding Fibonacci Hamiltonian~\eqref{eq:tight} with $T=6$ and $N=F_{16}=987$ sites. Top inset: 5 of the 987 eigenfunctions have energies that lie within the displayed energy window $W$. Bottom inset: three typical spatially decaying eigenfunctions, with nearly degenerate consecutive eigenvalues, as indicated within the inner panels.
}
\label{fig:energy_window}
\end{figure}

\begin{figure}
\centering
\includegraphics[width=\linewidth]{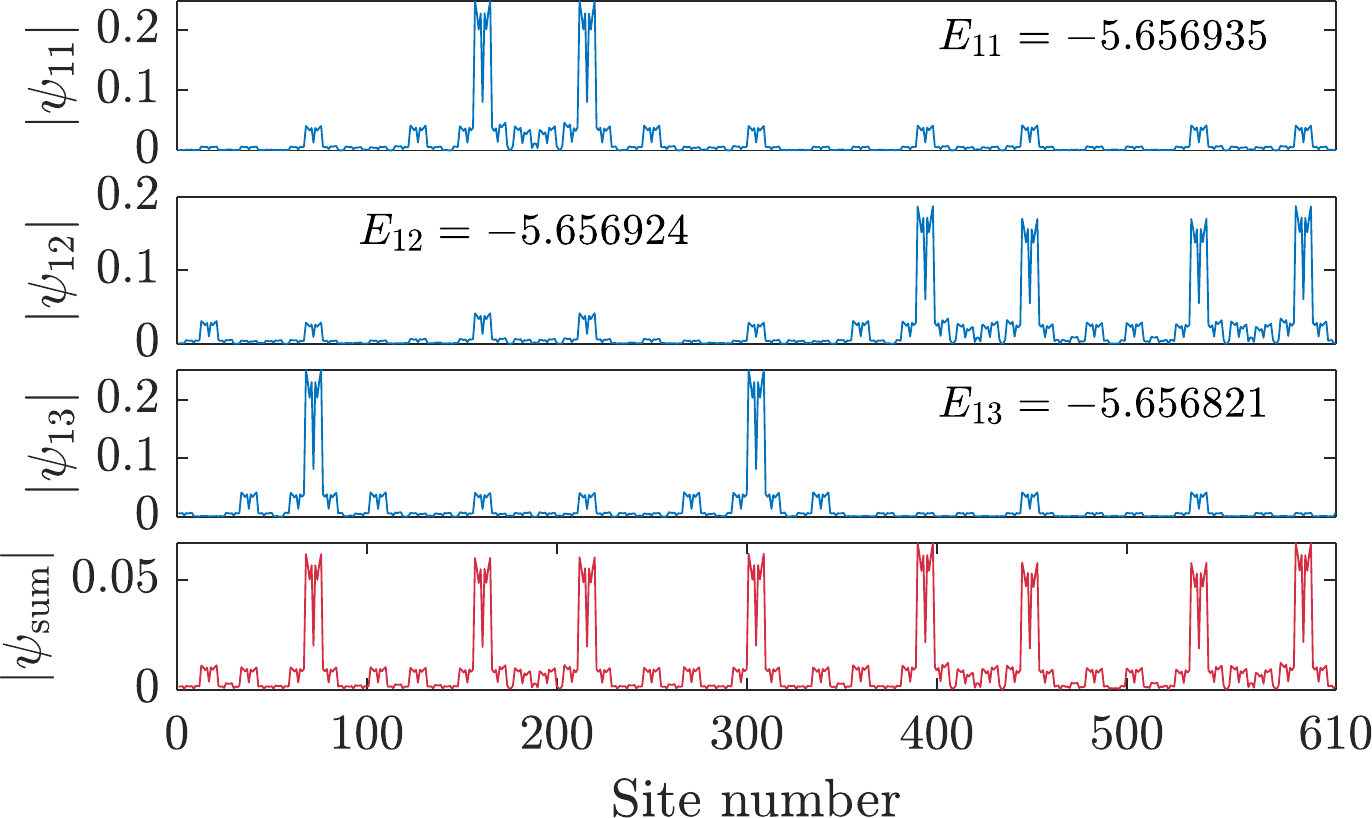}
\caption{The top three panels show three typical spatially decaying eigenfunctions of the Fibonacci quasicrystal, with nearly degenerate consecutive eigenvalues, as indicated within the panels, differing by less than $10^{-5}$ of the total spectral width. The bottom panel shows a linear combination of these three eigenfunctions, whose peaks are spatially extended, and for which the mean energy is $\langle E\rangle=-5.656905$, with an uncertainty, or standard deviation, of $\Delta E=4.2\cdot10^{-5}$. 
Here $T=6$ and $N=F_{15}=610$. }
\label{fig:lincomb_wave functions}
\end{figure}

The approximant tight-binding Hamiltonian, corresponding to Eq.~\eqref{eq:tight}, is thus given by a $F_n\cross F_n$ matrix that is tridiagonal up to boundary terms. A direct diagonalization of the matrix, for a given value of $T$ and a particular choice of boundary hopping terms, is then used to obtain a discrete set of $F_n$ eigenvalues along with their corresponding eigenfunctions, converging to the quasicrystalline spectrum in the limit of $n\to\infty$. A typical spectrum is shown in Fig.~\ref{fig:energy_window} for an approximant with $N=F_{16}=987$ sites, and $T=6$. Typical algebraically decaying eigenfunctions are shown in the bottom inset of Fig.~\ref{fig:energy_window}, as well as in blue in the top three panels of Fig.~\ref{fig:lincomb_wave functions}, for an approximant with $N=F_{15}=610$ sites, and $T=6$. Indeed, for any $T\neq1$ the eigenfunctions all tend in the infinite limit to critical wave functions that decay algebraically~\cite{Kohmoto83, ostlund, Niu86}, rather than to extended Bloch functions. 

\section{Linear combinations of nearly-degenerate eigenfunctions}

We wish to see whether extended Bloch wave functions may still appear naturally in these tight-binding models, albeit not as eigenfunctions. As a first step, and following early ideas of Even-Dar Mandel and Lifshitz~\cite{Shahar06, *Shahar08, ShaharThesis, ShaharArxiv}, we consider the spatial extent one can achieve by taking linear combinations of nearly degenerate eigenfunctions. This idea is motivated by the fact that many of the simple structural or tiling models of quasicrystals are \emph{linearly repetitive}. This means that any finite patch of radius $r$ is repeated in the tiling at a distance that scales linearly with $r$. Thus, algebraically decaying eigenfunctions that are very close in energy, which are likely to originate from large patches that are structurally similar, will have their peaks located at distant positions in the quasicrystal. This may allow a relatively small number of nearly degenerate eigenfunctions to span the whole quasicrystal. We note that \citet{Niu86} already saw indications that this might be the case by considering incoherent sums $\sum |\psi_n|^2$ of nearly degenerate eigenfunctions, but not as coherent wave functions $\sum c_n \psi_n$ that may describe an actual particle with a small uncertainly in its energy.

The three eigenfunctions shown in blue in the top three panels of Fig.~\ref{fig:lincomb_wave functions} correspond to three consecutive eigenenergies in the spectrum of a $N=F_{15}=610$ approximant, and together may indeed form a coherent superposition that spans the whole approximant, as shown in red in the bottom panel. The perceptive reader may notice that the peaks in the extended function are separated by long ($L$) and short ($S$) intervals, arranged according to the $N=F_{6}=8$ Fibonacci approximant $LSLLSLSL$. This reflects the spatial distribution of similar local sequences, in the $F_{15}$ approximant, inherited from its ancestor of 9 generations earlier.

\begin{figure}
\centering
\includegraphics[width=\linewidth]{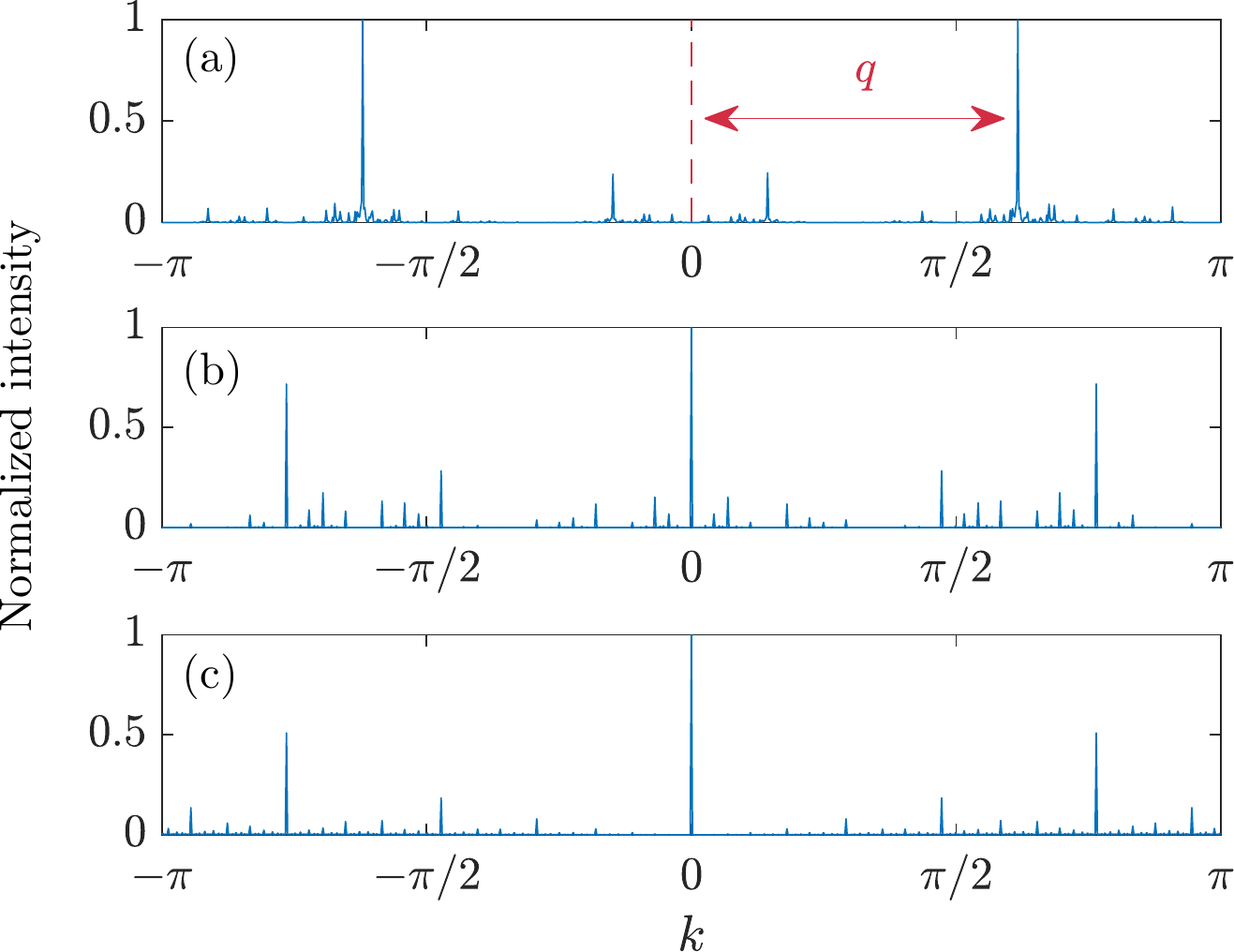}
\caption{Fourier spectra of (a)~$\psi$, (b)~$|\psi|$, and (c)~the potential, where $\psi$ is the most extended wave function generated by a linear combination of 34 nearly degenerate eigenfunctions of the Hamiltonian~\eqref{eq:tight}, calculated with $T=6$ and $N=F_{16}=987$ sites. The spectra of $|\psi|$ and the potential are peaked at the same wave vectors, while the spectrum of $\psi$ is shifted by a constant $q$, motivating us to label the wave function as $\psi_q$. The mean energy of the $\psi_q$ shown here is $E(q)=\bra{\psi_q}H\ket{\psi_q} = 0.1746$.}
\label{fig:finding_q}
\end{figure}

To pursue this intuition, we numerically optimize the spatial extent of linear combinations of nearly degenerate eigenfunctions, whose energies lie within tiny windows $W$ around every eigenenergy $E_n$ in the spectrum of a given approximant, as demonstrated schematically in the top inset of Fig.~\ref{fig:energy_window}. It is most often the case that several consecutive energies yield the same result owing to the near-flatness of the spectrum. We use the difference-map algorithm of \citet{Elser}, with the aim of generating a wave function that satisfies two constraints simultaneously: (I) it should be as extended as possible, ideally having an equal amplitude on all sites; and (II) it should be spanned by the assigned set of nearly degenerate eigenfunctions. The two constraints are applied iteratively---the first by setting $|\psi(m)|=1/\sqrt{N}$ on all sites, while keeping the phases, and the second by projecting the wave function into the subspace spanned by the given eigenfunctions---always finishing with the latter projection~\cite{ShaharArxiv, ShaharThesis}. We note that more traditional constraint-solving approaches, like least-squares algorithms, yield similar results, albeit with longer computation times. To assess the results and the progress of the calculation, we use the \emph{Normalized Participation Ratio},
\begin{equation}\label{eq:npr}
\NPR[\psi] = \frac{\left(\sum_m \left| \psi(m) \right| ^2 \right)^2}{N\sum_m \left| \psi(m) \right| ^4} \,,
\end{equation}
as a measure of extent. By construction, the NPR is equal to 1 for a uniformly extended wave function, and decays to 0 as it becomes localized.

\begin{figure}
\centering
\includegraphics[width=\linewidth]{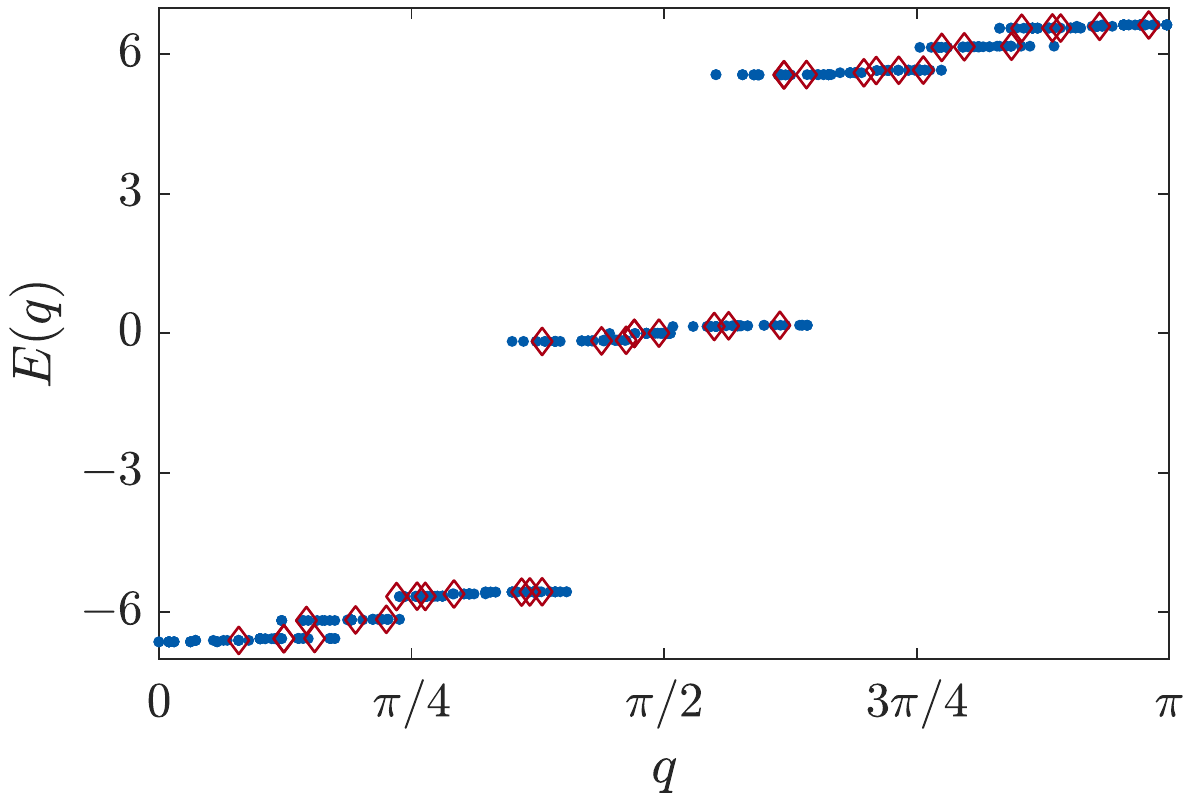}
\caption{Effective dispersion curve: mean energy $E$ of linear combinations of eigenfunctions, as a function of the extracted quantum number $q$. Red diamonds show the spectrum calculated from the most extended linear combinations of nearly degenerate eigenfunctions, yielding the most recent ancestor as a wave function. Blue dots include the addition of earlier ancestors. The calculation is performed with $T=6$, $N=987$ sites, and $W=0.005$, giving typically between 21 and 55 eigenfunctions in each linear combination. The dispersion curve resembles the shape of the energy spectrum in Fig.~\ref{fig:energy_window}, suggesting that $q \propto n$ as in periodic crystals.
}
\label{fig:E-q_dispersion}
\end{figure}

Surprisingly, the wave functions that emerge from this optimization all happen to be Bloch-like Fibonacci wave functions, even though they are merely optimized for extent. They are \emph{Fibonacci wave functions} in the sense that they have the structure of an earlier Fibonacci ancestor of the underlying potential, as anticipated in the bottom panel of Fig.~\ref{fig:lincomb_wave functions}. They are \emph{Bloch-like wave functions} in the sense that their Fourier spectra, as shown in Fig.~\ref{fig:finding_q}, are carried by the same set of wave-vectors as the underlying quasiperiodic potential---set by the Fibonacci sequence of hopping amplitudes $T_m$---shifted by the crystal momentum $q$. 

Although we are limited computationally up to approximants of size $N=F_{18}=2584$, it seems that this behavior persists with increasing $N$. 
Specifically, we find that while the NPRs of the individual eigenfunctions decrease with $N$, the NPRs of the most extended linear combinations of eigenfunctions remain roughly constant with increasing $N$, suggesting that the Bloch-like Fibonacci character of the extended linear combinations may persist in the thermodynamic limit.

Moreover, it is possible to extract the crystal momentum $q$ numerically for each energy, by comparing the Fourier transform of the potential to that of the extended wave function. Specifically, we calculate $q$ by the shift in the main, or strongest, peaks in the Fourier spectra of the wave function and the potential. We note that the accuracy in which we are able to determine $q$ improves with increasing approximant size. However, if the most extended wave function is a very early ancestor, as is the case in spectral regions with a low density of states, the resemblance between the Fourier spectra is weaker, and the determination of $q$ becomes less accurate.

The resulting effective dispersion relation $E(q)$, displayed as red diamonds in Fig.~\ref{fig:E-q_dispersion}, is very similar to the eigenenergy spectrum in Fig.~\ref{fig:energy_window}, suggesting that $q$ might be proportional to $n$. This is reminiscent of the situation for periodic crystals, where $q=2\pi n/N$, although some overlap exists between the apparent minibands, spoiling the perfect monotonic dependence of $E$ on $q$. We are unable to determine whether this overlap is real, or a result of not having obtained the optimal linear combination, leading to an inaccurate mapping of $q$ to $E$. 

Looking more quantitatively, we find that both the NPR and the ancestral generation, or Fibonacci number, associated with the most extended wave functions, are correlated with the spectral density of states around the mean energy of the extended wave function. As the density of states increases, a given fixed energy window $W$ will contain more eigenfunctions, allowing one to obtain a more extended wave function, and to more closely follow the Fibonacci potential, yielding a more recent ancestor. With values of $T$ large with respect to unity, as is the case for $T=6$, shown in Fig.~\ref{fig:energy_window}, the spectrum appears as a hierarchy of rather flat clusters of eigenvalues---each corresponding to a Fibonacci number of eigenfunctions---and the density of states is sharply peaked. In such cases a very small window $W$, of less than $10^{-3}$ of the full bandwidth, already contains a fairly large cluster of 21--55 of the 987 eigenfunctions, and a highly extended wave function corresponding to a recent ancestor is obtained. As $W$ decreases, fewer functions participate in the linear combination, the NPR of the optimized wave function gradually decreases, and the ancestral order increases, while the quantum number $q$ changes very little, although its numerical determination becomes less accurate. This qualitative behavior remains similar for all values of $T$. One should only note that as $T$ approaches unity, deviations from periodicity decrease, gaps in the spectrum gradually close, and the NPRs of the eigenfunctions themselves increase. 

\begin{figure}
\centering
\includegraphics[width=\columnwidth]{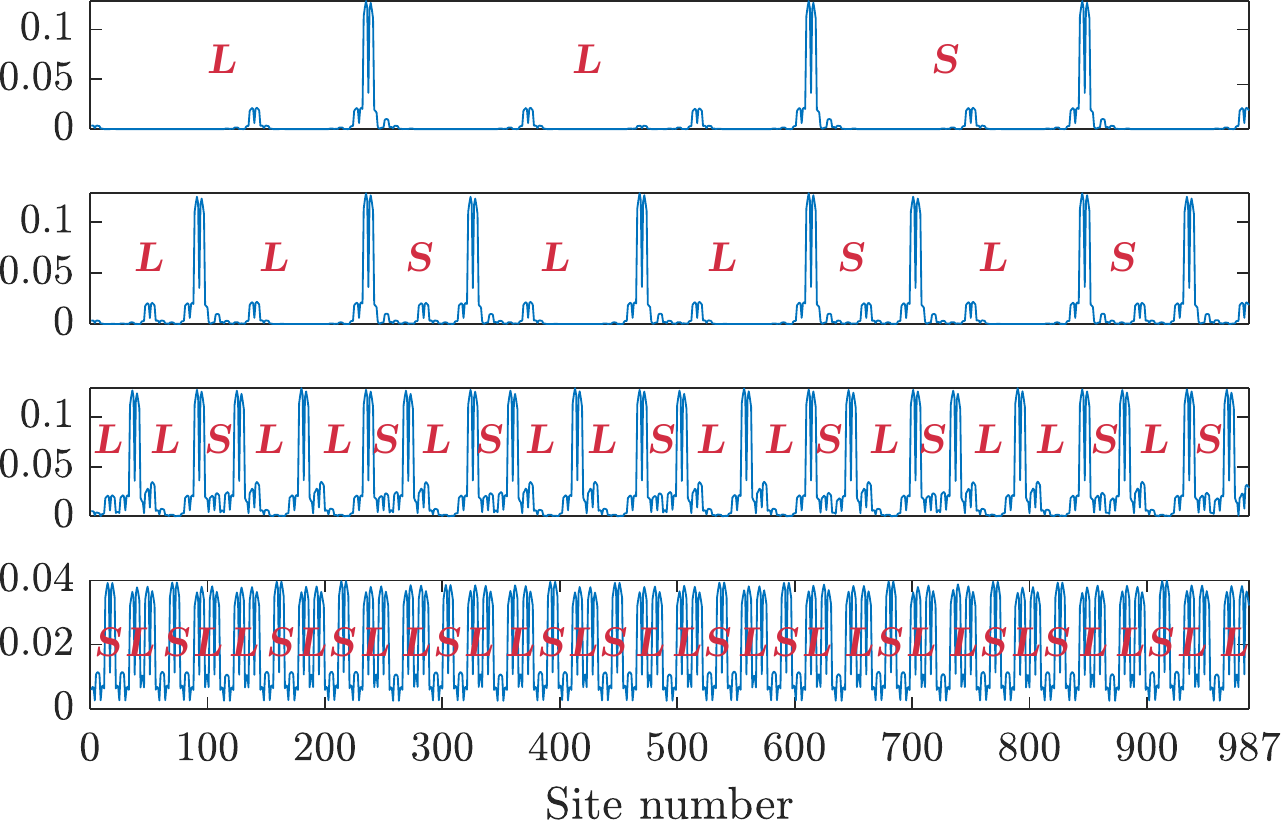}
\caption{Several early ancestors generated around the same eigenenergy $E_n=-6.6353$, calculated with $T=6$, $N=F_{16}=987$ sites, and an energy window $W=5\cdot10^{-3}$. The displayed ancestors, from top to bottom, are
$F_{4}=3$, with mean energy $\langle E\rangle=-6.632729$ and uncertainty $\Delta E=2.341\cdot10^{-3}$;
$F_{6}=8$, with $\langle E\rangle=-6.632730$, $\Delta E=2.341\cdot10^{-3}$;
$F_{8}=21$, with $\langle E\rangle=-6.632348$, $\Delta E=2.236\cdot10^{-3}$;
and the optimal ancestor $F_{9}=34$, with $\langle E\rangle=-6.635101$ and $\Delta E=8.382\cdot10^{-4}$.
}
\label{fig:all_ancestors}
\end{figure}

Rather than optimizing for extent, we can employ standard distance-minimization algorithms using earlier ancestors as targets for optimization, without decreasing the number of eigenfunctions used. 
The result of such a construction is shown in Fig.~\ref{fig:all_ancestors}, where four different ancestors are obtained as linear combinations of the same set of nearly degenerate eigenfunctions. We find that these earlier ancestors have similar values of $q$, which change monotonically with their mean energies. This seems to imply that earlier ancestors provide additional samples of the dispersion relation $E(q)=\bra{\psi_q}H\ket{\psi_q}$, generating a denser dispersion curve, shown as blue dots in Fig.~\ref{fig:E-q_dispersion}. However, this curve exhibits more overlap between the minibands, with $q\propto n$ only within each miniband, perhaps because the determination of $q$ becomes less accurate the earlier the ancestor.

\section{Addition of weak disorder}

Our analysis indicates that a small uncertainty $\Delta E \simeq W$, in the energy of a single electron, may allow it to explore sufficiently many nearly degenerate energy eigenfunctions for it to behave like a Bloch electron. Such an uncertainty may arise naturally from semiclassical dynamics in a ``semiadiabatic'' regime~\cite{Spurrier18} that smooths out small gaps in the spectrum, leaving effective minibands $E(q)$ between the remaining large gaps. Alternatively, as we explore here, a mixing of nearly degenerate eigenfunctions may be induced by adding weak disorder to the otherwise perfect quasicrystalline Hamiltonian~\eqref{eq:tight}. We model the disorder using zero-mean Gaussian random variables with standard deviation $\Delta T$, added to the hopping amplitudes $T_m$ in Eq.~\eqref{eq:tight}. Similar results are obtained for $\delta$-correlated, or white, disorder as for inverse power-law correlated disorder, so we focus here on the former. We perform an ensemble average of typically 1000 disorder realizations for each eigenfunction. The disorder is kept sufficiently weak to maintain monotonic dependence of the energy eigenvalues on disorder strength, so as not to change the identity of the eigenfunctions by approaching level crossings. 

\begin{figure}
\centering
\includegraphics[width=\linewidth]{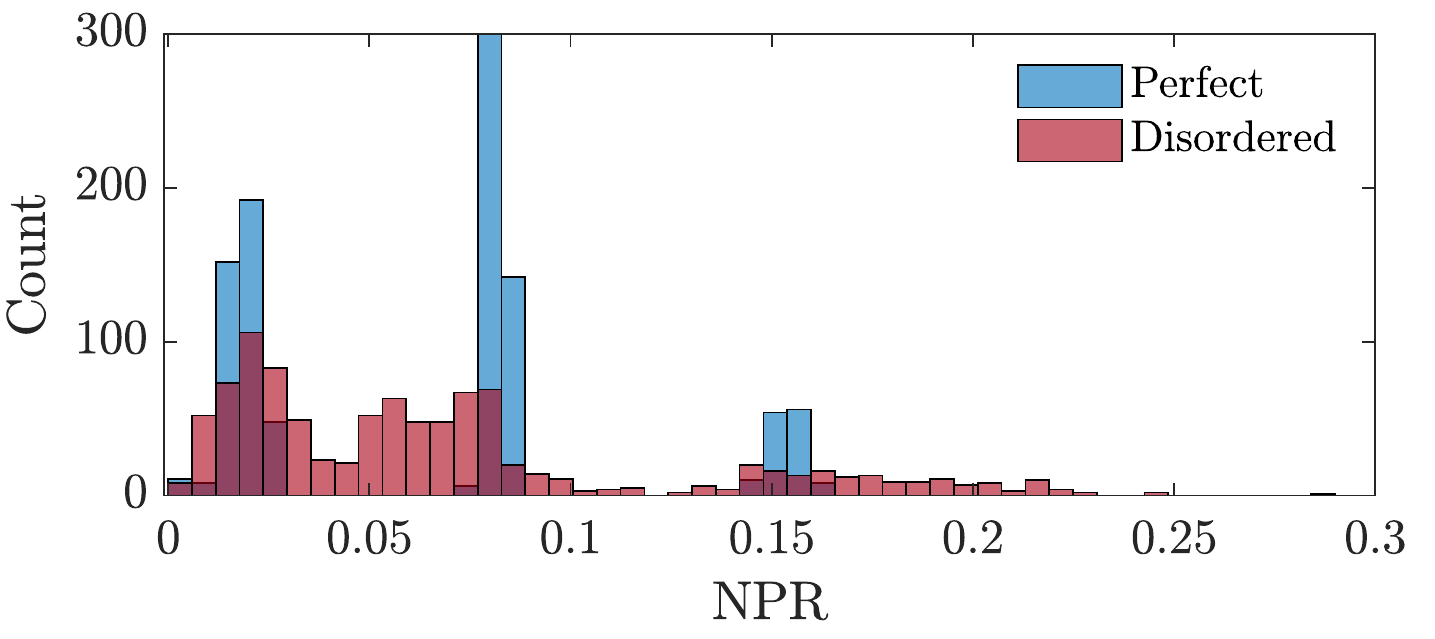}
\caption{Histograms of the NPR of the eigenfunctions of the perfect Fibonacci Hamiltonian~\eqref{eq:tight} in blue, and those of the Hamiltonian with added white Gaussian disorder of strength $\Delta T=10^{-4}$, each averaged over $1000$ disorder realizations, in red, calculated with $T=6$ and $N=987$. Weak disorder increases the NPR of the most extended unperturbed eigenfunctions and distributes the NPR values more evenly across the histogram.}
\label{fig:part_orig_vs_noisy} 
\end{figure}

As expected, extremely weak disorder hardly affects the eigenfunctions, while sufficiently large disorder leads to Anderson localization, \emph{i.e.}\ to exponentially decaying eigenfunctions. A little less expected, yet consistent with previous observations~\cite{Levi11,Roche97,Jagannathan19}, is the result that weak disorder, on the order of the windows $W$ used earlier or less, improves the extent of some of the eigenfunctions, while reducing the extent of others. This is demonstrated in the histograms of Fig.~\ref{fig:part_orig_vs_noisy}, where one can see that the sharply distributed NPR values in the perfect Hamiltonian are more evenly distributed in the ensemble-averaged disordered Hamiltonian. We note that for the disorder strength giving rise to these results, the energy spectrum itself remains nearly unchanged.

\begin{figure}[t]
\centering
\includegraphics[width=\linewidth]{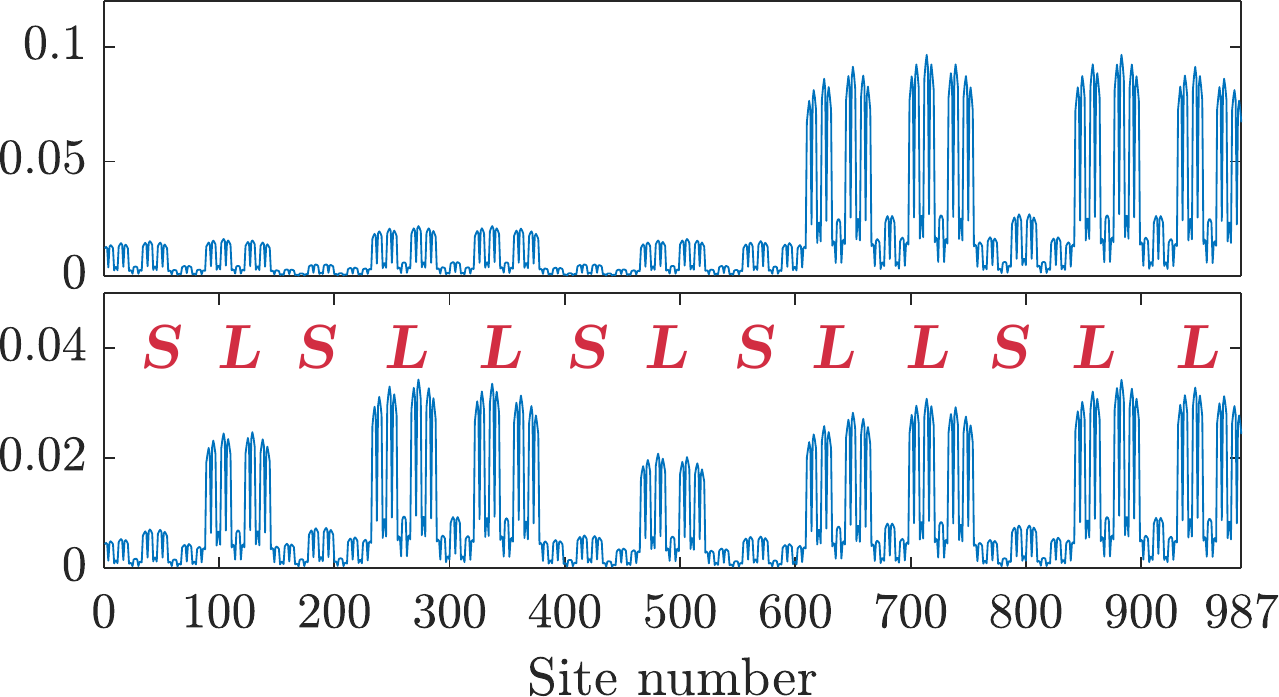}
\caption{Comparison between an eigenfunction of the perfect Fibonacci quasicrystal with $T=6$ and $N=F_{16}=987$ sites (top), and the ensemble average of 1000 realizations of the corresponding eigenfunction of the disordered Hamiltonian with $\Delta T=10^{-4}$ (bottom). Here, disorder increases the NPR from $0.16$ to $0.28$, and the ensemble-averaged eigenfunction has the structure of a nearly perfect $N=F_{7}=13$ Fibonacci ancestor.}
\label{fig:noise_wave functions} 
\end{figure}

More important is the observation that weak disorder significantly increases the NPR of the most extended eigenfunctions, and that the resulting ensemble-averaged functions take the form of nearly perfect early Fibonacci ancestors, like the ones obtained above by optimizing linear combinations of eigenfunctions of the perfect Hamiltonian. However, since these are very early Fibonacci ancestors, we are unable to faithfully determine their corresponding wave vector $q$. The most extended eigenfunctions, obtained with added disorder of a given strength, are quite similar to each other and resemble the Fibonacci ancestor shown in Fig.~\ref{fig:noise_wave functions}. 

\section{Conclusion}

The study of quasicrystals has taught us that the electronic properties of materials depend crucially, not only on \emph{how ordered they are}, but also on \emph{how they are ordered}. As we demonstrate here, the interplay between disorder and aperiodic long-range order is particularly intriguing. 

Owing to the dense nature of the Fourier module $\CL$, the Bloch sum over wave vectors in Eq.~\eqref{Eq:Bloch} is not guaranteed to behave properly for quasicrystals, as it does for periodic crystals. As a consequence, Bloch's theorem generally fails and eigenfunctions are generically critical, decaying algebraically rather than being quasiperiodically extended throughout the crystal. Nevertheless, quasiperiodic Bloch wave functions do form as superpositions of relatively small numbers of nearly degenerate critical eigenfunctions, which emerge naturally in the presence of weak disorder. Contrary to its effect on periodic crystals, disorder in quasicrystals first increases the extent of the most extended eigenfunctions, transforming them from critical to extended quasiperiodic Bloch functions, before eventually giving way to Anderson's inevitable localization. 

As for periodic crystals, one can associate an effective crystal momentum $q$ with each quasiperiodic Bloch wave function that forms, leading to an effective dispersion relation $E(q)$, which may explain the dispersion curves observed in certain experiments~\cite{rotenberg,Chiang19}. We expect this behavior to occur in other linearly repetitive quasicrystalline models, and to be even more pronounced in two and in three dimensions, where Anderson localization is less restrictive. We leave the verification of this expectation, as well as the analytical explanation of our empirical findings, and their extension to other models---perhaps considering models like the Thue-Morse sequence that are deterministic yet possess no long-range order---for future research.

\begin {acknowledgments}
The authors are grateful to Shahar Even-Dar Mandel for his helpful advice. RL thanks Alastair Rucklidge for fruitful and engaging discussions while on leave at the University of Leeds, where this manuscript was finalized. RL also thanks the School of Mathematics at the University of Leeds for their kind hospitality, and the Cheney Foundation for their financial support. Funding for this research was provided by the Israel Science Foundation (grant No.~1667/16).
\end {acknowledgments}

\bibliography{Bloch}

\end{document}